\documentclass[onecolumn,nobibnotes,nofootinbib]{revtex4}
\usepackage{amsmath,amssymb,bm}
\usepackage{epsfig}
\usepackage{graphicx}
\usepackage{pstricks,pst-node,pst-text,pst-3d}

\newcommand{\avg}[1]{\left\langle #1 \right\rangle}
\newcommand{\vecr}{\bm{r}}
\newcommand{\vecb}{\bm{b}}

\newcommand{\x}{\bm{x}}
\newcommand{\y}{\bm{y}}
\newcommand{\z}{\bm{z}}

\begin{document}
\title{\bf DIS and the effects of fluctuations: a momentum space analysis}
 \author{E. Basso}
  \email{ebasso@if.ufrgs.br}
 
 \author{M. B. Gay Ducati}
  \email{beatriz.gay@ufrgs.br}
 
 \author{E. G. de Oliveira}
  \email{emmanuel.deoliveira@ufrgs.br}
 
  \author{J. T. de Santana Amaral}
  \email{thiago.amaral@ufrgs.br}
  \affiliation{Instituto de F\'{\i}sica, Universidade Federal do Rio
                        Grande do Sul, Caixa Postal 15051, 91501-970 - Porto Alegre, RS,
    Brazil}



\begin{abstract}
Among the dipole models of deep inelastic scattering at small
values of the Bjorken variable $x$, one has been recently
proposed which relates the virtual photon-proton cross
section to the dipole-proton forward scattering amplitude
in momentum space. The latter is parametrized by an expression
which interpolates between its behavior at saturation and
the travelling wave, ul\-tra\-vio\-let, amplitudes predicted by
perturbative QCD from the Balitsky-Kovchegov equation.
Inspired by recent developments in coordinate space, we
use this model to parametrize the proton structure function
and confront it to HERA data on $ep$ deep inelastic scattering.
Both event-by-event and the physical amplitudes are considered,
the latter used to investigate the effect of gluon number fluctuations, beyond
the mean-field approximation. We conclude that fluctuations
are not present in DIS at HERA energies.


\end{abstract}

\maketitle

\section{Introduction}

The well known correspondence between the evolution in QCD at small-$x$ (at high energy)
and a reaction-diffusion process has been the main source of the recent
knowledge concerning the scattering amplitudes and their evolution towards
the high energy limit. In particular, it has been realized that
the Balitsky-JIMWLK equations do not take into account the influence of
the \textit{gluon number fluctuations} \cite{IMM04,IT04,IM032,MS04}. In the mean field approximation,
these equations reduce to a unique equation, the so-called Balitsky-Kovchegov
(BK) equation \cite{bal,kov}, the simplest equation that describes the
evolution of the dipole scattering amplitude with rapidity $Y\equiv\ln 1/x$.
Being a mean field version of Balitsky-JIMWLK equations, BK equation does not
include the effects of discreteness and consequently of the fluctuations.
Among the consequences of the fluctuations, at least in the fixed coupling case,
one can cite, for example, the slowing down of the approach towards
the unitarity limit, as compared to the mean field framework, and the break down
of the \textit{geometric scaling} \cite{gscaling}, a phenomenological feature
observed at the DESY $ep$ collider HERA, in the measurements of inclusive
$\gamma^*p$ scattering, which is naturally explained in terms of the traveling
wave solutions of BK equation \cite{mp}.

At small-$x$, electron-proton deep inelastic scattering (DIS) can be seen in a particular
frame, called \textit{dipole frame}, which allows the factorization of the virtual photon-proton
cross section. In this frame, the proton carries most of the total energy,
but the photon has enough energy to split into a quark-antiquark pair, or a
dipole. This $q\bar{q}$ pair then interacts with the proton. The virtual
photon-proton cross section can be written as

\begin{equation}\label{eq:cross_section}
  \sigma_{T,L}^{\gamma^{*}p}(Q^2,Y)=\int
  d^2r\int_{0}^{1}dz\,\left|\Psi_{T,L}(\bm{r},z;Q^2)\right|^{2}
  \sigma_{\text{dip}}(\bm{r},Y),
\end{equation}
where the labels $T$ and $L$ refer, respectively, to the transverse and
longitudinal parts of the cross section, $\bm{r=x-y}$ is the vector which
gives the transverse size of the dipole, $\bm{x}$ and $\bm{y}$ being the
transverse coordinates of the quark and the antiquark; $z$ is the fraction
of the momentum of the photon carried by the quark and $\Psi_{T,L}(\bm{r},z;Q^2)$
are the transverse and longitudinal wave functions for the photon to go into
the dipole, whose explicit forms are well known from QED. Using
Eq.(\ref{eq:cross_section}) one can obtain the expression for the $F_2$ proton
structure function through the formula
\begin{eqnarray}
F_2(x,Q^2)=\frac{Q^2}{4\pi^2\alpha_{em}}
  \left[\sigma_T^{\gamma^*p}(x,Q^2)+\sigma_L^{\gamma^*p}(x,Q^2)\right].
\end{eqnarray}

The quantity $\sigma_{\textrm{dip}}$ in Eq.(\ref{eq:cross_section}) is the dipole-proton
cross section which can be expressed as

\begin{equation}\label{eq:sigma-dip}
\sigma_{dip}^{\gamma^{*}p}(\bm{r},Y)=2\int d^2\bm{b}\,\avg{T(\bm{r},\bm{b})}_Y,
\end{equation}
where $\avg{T(\bm{r},\bm{b})}_Y$ is the average  scattering amplitude -- the
notation $\avg{\cdot}$ denotes the average over all the realizations of
the target (proton) color field -- for the dipole-proton scattering at a
given impact parameter $\bm{b}=(\bm{x}+\bm{y})/2$.


As the explicit forms of the virtual photon wave functions are well known,
one is left with the parametrization of the scattering amplitude, that is,
one has to model the dipole-proton cross section. Different approaches have
already been proven successful, giving good description of the data, among
them one can cite the model proposed by Golec-Biernat and Wusthoff,
the GBW saturation model \cite{gbw}, and the model by Iancu, Itakura and Munier,
the IIM model or CGC fit \cite{iim}, both developed in coordinate
space. These models have been recently considered in order to study the 
effects of fluctuations on DIS in the fixed coupling case \cite{Kozlov:2007wm}.
Although the description of DIS data is improved once gluon number fluctuations
are included, it is not possible to state that the improvements come from
them, but they may come from the geometric scaling violation present, for example,
in the diffusion part of BK solution present in the IIM model.

In this work we perform an analysis similar to that done in \cite{Kozlov:2007wm} by
using a recently proposed parametrization for the dipole-proton scattering amplitude,
the first developed in momentum space \cite{agbs}.

In Section II we present a review on the main features of the QCD evolution
at high energies within the dipole picture \cite{dipolepic}. Section
III is devoted to describe the parametrization for the scattering amplitude
in momentum space that will be used to describe the DIS data. The fluctuations
are properly included, the average amplitude is defined and the results of
the fit to the last HERA data are presented. The discussion and conclusions
are presented in Section IV.

\section{The dipole scattering amplitude}

\subsection{Dipole evolution}
Considering a dipole of transverse coordinates $\bm{x}$ and $\bm{y}$ at a given
rapidity $Y$, if one increases the rapidity from $Y$ to $Y+\delta Y$, there
is a probability for a gluon, with transverse coordinate $\bm{z}$, to be
emitted by the quark (or antiquark) of the pair. In the large $N_c$ limit
($N_c$ is the number of colors), this gluon can be considered as a quark-antiquark
pair --a new dipole--  at point $\bm{z}$. This is the dipole picture introduced by
Mueller \cite{dipolepic}.

The probability density for the original dipole to split into the two child dipoles
($\bm{x},\bm{z}$) and ($\bm{z},\bm{y}$) is given by
\begin{equation}
\frac{\bar{\alpha}}{2\pi}{\cal M}(\bm{x},\bm{y},\bm{z})\,dY\,d^2z,
\end{equation}
where $\bar{\alpha}=\alpha_sN_c/\pi$ and
 \begin{equation}
{\cal M}(\x,\y,\z)= \frac{(\x-\y)^2}{(\x-\z)^2(\z-\y)^2}.
\end{equation}
If the target (in this case, the proton) is dense enough, both child dipoles can
interact with it and the resulting evolution equation for the dipole scattering
amplitude is \cite{bal}

\begin{equation}\label{eq:b-jimwlk}
\partial_Y \avg{T(\x,\y)} = \bar{\alpha}\int d^2z\,{\cal M}(\x,\y,\z)
\left[\avg{T(\x,\z)}+\avg{T(\z,\y)}-\avg{T(\x,\y)}
-\avg{T(\x,\z)T(\z,\y)}\right].
\end{equation}
This is not a closed equation for the one-dipole scattering amplitude, but the first
equation of an infinite hierarchy, the Balitsky-JIMWLK hierarchy \cite{bal,jkmw97}. The
first and second terms give the contribution of each dipole which can interact with the
proton; the third term gives the virtual contributions to the scattering and the last
suppression (quadratic) term comes from taking into account multiple interactions, when
both child dipoles interact with the target. When $T$ is small, this quadratic term can
be neglected and Eq.(\ref{eq:b-jimwlk}) reduces to the (dipole version of) the linear
BFKL equation \cite{bfkl}.

\subsection{Balitsky-Kovchegov equation}

In the mean field approximation, valid when the target is sufficiently large and
homogeneous, one can write $\avg{T(\x,\z)T(\z,\y)}\approx \avg{T(\x,\z)}\avg{T(\z,\y)}$
and the resulting equation is the so-called BK (Balitsky-Kovchegov) equation \cite{bal,kov}

\begin{equation}\label{bk}
\partial_Y \avg{T(\x,\y)} = \bar{\alpha}\int d^2z\,{\cal M}(\x,\y,\z)\left[\avg{T(\x,\z)}
+\avg{T(\z,\y)}-\avg{T(\x,\y)} -\avg{T(\x,\z)}\avg{T(\z,\y)} \right].
 \end{equation}
This equation includes unitarity corrections and is free from the
problem of diffusion to the infrared (nonperturbative) region, present in the
solution of BFKL equation, since there is a scale, the saturation scale $Q_s(Y)$, an increasing function of rapidity, which naturally emerges from BK equation.

If one neglects the dependence on the impact parameter, the scattering amplitude
depends only on the size of the dipole and Eq.(\ref{bk}) becomes an equation for
$\avg{T(r)}$, where $r=|\vecr|=|\x-\y|$. Let us denote this mean-field amplitude
by ${\cal N}_Y(r)$. After performing the Fourier transform

\begin{equation}\label{eq:fourier}
 N_Y(k)=\frac{1}{2\pi} \int \frac{d^2r}{r^2}\,e^{i\mathbf{k}\cdot\mathbf{r}}\,{\cal N}_Y(r) = \int_0^\infty\frac{dr}{r}J_0(kr){\cal N}_Y(r),
\end{equation}
one gets the $\vecb$-independent BK equation in momentum space
\begin{equation}
\partial_Y N_Y(k) = \frac{\bar\alpha}{\pi}\int \frac{dp^2}{p^2}
                    \left[\frac{p^2N_Y(p)-k^2N_Y(k)}{|k^2-p^2|} + \frac{k^2N_Y(k)}{\sqrt{4p^4+k^4}}\right]
                  - \bar\alpha N_Y^2(k),
\end{equation}
which can be rewritten as
\begin{equation}\label{eq:bk}
\partial_Y N_Y=\bar{\alpha}\chi(-\partial_L)N_Y-\bar{\alpha}N_Y^{2},
\end{equation}
where
\begin{equation}\label{eq:kernel}
\chi(\gamma)=2\psi(1)-\psi(\gamma)-\psi(1-\gamma)
\end{equation}
is the BFKL kernel and $L=\log(k^2/k_0^2)$, with  $k_0$ some fixed soft scale.

The  kernel (\ref{eq:kernel}) can be written with the help of a series and after
a saddle point approximation and a change of variables \cite{mp} it has been shown
that BK equation reduces to Fisher and Kolmogorov-Petrovsky-Piscounov  (FKPP)
equation \cite{fkpp}, which admits the so-called traveling wave solutions.
In the QCD language this means that, at asymptotic rapidities, the scattering amplitude
depends only on a single variable, the ratio $k^2/Q_s^2(Y)$ instead of depending
separately on $k^2$ and $Y$. This scaling property is called geometric scaling
and has been observed in the measurements of the proton structure function at HERA \cite{gscaling}.
The amplitude is a wavefront which interpolates between 0 and 1 and travels towards large
values of $k^2$ with speed $\lambda$ -- the saturation exponent -- without deformation.
The position of the front, for which  $N_Y={\cal O}(1)$ is given by the saturation
momentum $Q_s(Y)$ or, more specifically, by $\log (Q_s^2(Y)/k_0^2)=\lambda Y$.

The expression for the tail of the scattering amplitude reads
\begin{equation}\label{eq:Ttail}
N_Y\left(k\right) \stackrel{k\gg Q_s}{\approx}
  \left(\frac{k^2}{Q_s^2(Y)}\right)^{-\gamma_c}\log\left(\frac{k^2}{Q_s^2(Y)}\right)
\exp\left[-\frac{\log^2\left(k^2/Q_s^2(Y)\right)}{2\bar{\alpha}\chi''(\gamma_c)Y}\right],
\end{equation}
where
\begin{equation}\label{eq:lambda}
\lambda =\min\, \bar{\alpha}\frac{\chi(\gamma)}{\gamma}=\bar{\alpha}\frac{\chi(\gamma_c)}{\gamma_c}=\bar{\alpha}\chi^\prime(\gamma_c).
\end{equation}
One can observe that the last term in Eq.(\ref{eq:Ttail}) has an important role, since it
introduces an explicit dependence on rapidity and hence violates geometric scaling.
Then, geometric scaling is obtained for
\begin{equation}
\log\left(k^2/Q_s^2(Y)\right) \lesssim \sqrt{2\chi''(\gamma_c)\bar{\alpha} Y},
\end{equation}
i.e., within a window $\sqrt{Y}$ above the saturation scale.

\subsection{Beyond the mean field approximation: the effects of fluctuations}

The BK equation is the simplest equation which describes high energy dipole
evolution and scattering in perturbative QCD. This mean field equation has been shown to be in the
universality class of FKPP equation, whose dynamics is called reaction-diffusion
dynamics. Within this correspondence between a reaction-diffusion process and
the QCD evolution at high energy, it has been recently realized that the Balitsky-JIMWLK
hierarchy is not complete because they do not take into account the gluon (dipoles) number
fluctuations, which are related to discreteness in the evolution, and thus
they are completely missed by BK equation.

As we will see in the following, the fluctuations influence dramatically the
QCD evolution at high energies, and so the properties of the scattering
amplitudes. Their inclusion results in a new hierarchy of evolution equations,
the \textit{Pomeron loop} equations \cite{IT04}. The first equation of this hierarchy
is exactly the same as Eq.(\ref{eq:b-jimwlk}), but as one goes to the second one, the
evolution equation for the two-dipole amplitude $\avg{T^{(2)}}\equiv \avg{TT}$, besides
the linear (BFKL) term and the nonlinear term, responsible for unitarity
corrections, there is a new term, proportional to the one-dipole amplitude
$\avg{T}$, which is the fluctuation term. More generally, the equation for the
$k$-dipole amplitude $\avg{T^{(k)}}$ depends on $\avg{T^{(k)}}$, $\avg{T^{(k+1)}}$ and $\avg{T^{(k-1)}}$, the
latter being the contribution of the fluctuation. After an approximation
\cite{IT04} to get rid of the impact-parameter dependence, this can
be written as a Langevin equation for the event-by-event amplitude
which is formally the BK equation with a noise term, which lies in the
same universality class of the stochastic FKPP equation (sFKPP).

Each realization of the noise means a single realization of the target
in the evolution and leads to an amplitude for a single event.
Different realizations of the target lead to a dispersion of the
solutions, and then in the saturation momentum $\rho_s\equiv \ln(Q_s^2/k_0^2)$
from one event to another. The saturation scale is now a random variable whose average value is given by
\begin{equation}
\langle Q_s^2(Y) \rangle = \exp{[\lambda^*Y]}.
\end{equation}

The dispersion in the position of the individual fronts is given by
\begin{equation}
\sigma^2 = \langle \rho_s^2 \rangle - \langle \rho_s \rangle^2 = D\bar{\alpha}Y.
\end{equation}
The \textit{diffusion coefficient} $D$, as well as the average saturation exponent
$\lambda^*$, are analytically known only in the asymptotic limit $\alpha_s^2 \rightarrow 0$, then in what follows they will be treated as free parameters.

The probability distribution of $\rho_s$ is, to a good approximation, a Gaussian \cite{Marquet:2006xm}
\begin{equation}
P_Y(\rho_s)\simeq \frac{1}{\sqrt{\pi\sigma^2}}\exp\left[-\frac{(\rho_s-\avg{\rho_s})^2}{\sigma^2}\right].
\end{equation}
For each single event, the evolved amplitude shows a traveling-wave
pattern, which means that geometric scaling is preserved for each
realization of the noise. However, the speed of the wave is smaller
than the speed predicted by BK 	equation. This speed, or the (average)
saturation exponent, has been found to be \cite{IT04}
\begin{equation}\label{eq:speed-fluct}
\lambda^* \simeq \lambda -\frac{\pi^2\gamma_c\chi^{\prime\prime}(\gamma_c)}{\ln(1/\alpha_s^2)}.
\end{equation}
The average amplitude is determined by ($X\equiv \ln(1/r^2Q_0^2)$)
\begin{equation}
\avg{T(X,\rho_s)} = \int^{+\infty}_{-\infty}d\rho_s\,P_Y(\rho_s)T(X,\rho_s).
\end{equation}
A crucial property of the physical amplitudes is that at sufficiently high energies,
unlike the individual fronts, they will generally not show geometric scaling. More
specifically, they  will show additional dependencies upon $Y$, through the front dispersion
$\sigma$. Then, geometric scaling is washed out and replaced by the so-called \textit{diffusive
scaling} \cite{IT04,IM032,MS04,Hat06}
\begin{equation}\label{eq:avg-amplitude}
\avg{T(X,\rho_s)} = {\cal T}\left(\frac{X-\avg{\rho_s}}{\sqrt{\bar{\alpha}DY}}\right).
\end{equation}

\section{Description of DIS data}

We have seen in the Introduction that in the dipole frame the $F_2$ proton structure function
can be written in terms of the dipole-proton cross section $\sigma_{dip}$, which
can be expressed through the average dipole-proton scattering amplitude through
Eq.(\ref{eq:sigma-dip}). If one treats the proton as an homogeneous disk of radius
$R_p$, i.e., if one neglects the impact parameter dependence, the amplitude depends
only upon the size $r$ of the dipole and after integrating out the remaining angular
dependence, the dipole-proton cross section can be written in terms of the amplitude
$\avg{T(r)}$ in the following way:
\begin{equation}\label{eq:sigmadip-r}
\sigma_{dip}^{\gamma^{*}p}(r,Y)=2\pi R_{p}^{2}\avg{T(r)}.
\end{equation}
This expression must be inserted in Eq.(\ref{eq:cross_section}) and one has to
parametrize the dipole scattering amplitude $\avg{T(r)}$ in order to reproduce
$\sigma^{\gamma^*p}$ measurements. From now on we will denote this amplitude
by $T(r,Y)$.

\subsection{AGBS model and fluctuations}

The most recent parametrization for the dipole-proton scattering amplitude
has been proposed by Amaral, Gay Ducati, Betemps and Soyez, the AGBS model
\cite{agbs}, which is the first parametrization in momentum space in the literature
and is based on the knowledge of asymptotic behaviors of the solutions of BK equation.
The starting point is that, after performing the Fourier transform
(\ref{eq:fourier}), it is possible to rewrite the cross section (\ref{eq:cross_section})
in terms of the amplitude in momentum space. The $F_2$ structure function takes the form \cite{agbs}
\begin{equation}\label{eq:f2-mom}
F_2(x,Q^2)=\frac{Q^2R_p^2N_c}{4\pi^2}\int_0^\infty\frac{dk}{k}\int_0^1 dz\,
|\tilde{\Psi}(k^2,z;Q^2)|^2 \tilde{T}(k,Y),
\end{equation}
where now the photon wave function is expressed in momentum space and $\tilde{T}(k,Y)$ is the
scattering amplitude in momentum space. The AGBS model
analytically interpolates between the behaviors of the BK scattering amplitude
in the dilute regime, which is described by Eq.(\ref{eq:Ttail}), and the saturation
one, in which it behaves like
\begin{equation}\label{eq:Tsat}
\tilde{T}(k,Y)\left(k\right) \stackrel{k\ll Q_s}{=} c-\log\left(\frac{k}{Q_s(Y)}\right).
\end{equation}
If one defines the variable $\rho\equiv \ln (k^2/k_0^2)$, the interpolation in the
AGBS model is done through the following expression for the scattering amplitude
\begin{equation}\label{eq:Tmodel}
\tilde{T}^{\rm{AGBS}}(\rho,Y) = L_F \, \left(1-e^{-T_{\text{dil}}}\right),
\end{equation}
where
\begin{equation}\label{eq:Tdil}
T_{\text{dil}} = \exp\left[-\gamma_c\left(\rho-\rho_s\right)
-\frac{{\cal L}^2-\log^2(2)}{2\bar{\alpha}\chi''(\gamma_c)Y}\right],
\end{equation}
\begin{equation}\label{eq:l_red}
{\cal L}=
\ln \left[1+e^{(\rho-\rho_s)}\right] \qquad \textrm{with} \quad Q_s^2(Y) = k_0^2\,e^{\lambda Y},
\end{equation}
and
\begin{equation}
L_F=
1+\ln \left[e^{\frac{1}{2}(\rho-\rho_s)}+e^{-\frac{1}{2}(\rho-\rho_s)}\right].
\end{equation}
Through this parametrization, the measurements for the $F_2$ structure function were successfully reproduced with the contributions of light and heavy (charm) quarks included in the fit.


To include the fluctuations in the description of HERA data, one considers
the scattering amplitude given by AGBS model as a single event one, and one has
to evaluate the average scattering amplitude, which is obtained by performing the
integration
\begin{equation}\label{eq:agbs-fluct}
\avg{\tilde{T}_Y^{\rm{AGBS}}(\rho,\avg{\rho_s})}=\int^{+\infty}_{-\infty}d\rho_s\,P_Y(\rho_s)\tilde{T}_Y^{\rm{AGBS}}(\rho,\rho_s).
\end{equation}

This is the expression which must be inserted into (\ref{eq:f2-mom})
in order to reproduce DIS measurements of $F_2$ structure function
with fluctuations effects correctly included.


\section{Data set and results}
In this analysis, all the last HERA data measurements of the proton structure
function from H1 and ZEUS Collaborations \cite{h1,zeus} are fitted, within the
following kinematical range:
\begin{equation}
x\leq 0.01,
\end{equation}
\begin{equation}
0.045\leq Q^2 \leq 150\,\rm{ GeV}^2,
\end{equation}
which corresponds to 279 data points. Both ranges include values of $x$ low enough for the
analysis to be in the high energy regime, and values of $Q^2$ which allow us not to include
DGLAP corrections.

Concerning the parameters, we keep fixed $\bar{\alpha}=0.2$, which enters into the
amplitude through Eq.(\ref{eq:lambda}), and $\gamma_c=0.6275$, whose value corresponds
to the one obtained from the LO BFKL kernel. The other parameters in the amplitude,
$\lambda$, $k_0^2$ and $\chi^{\prime\prime}$, are left to be free, as well as the
proton radius $R_p$, which fixes the normalization of the dipole-proton cross section
with respect to the dipole-proton amplitude, and the diffusion coefficient $D$.
Only light quarks are considered and the values used for their masses are
$m_{u,d,s}=50$ and $140$ MeV.

\begin{table}
\begin{tabular}{l@{\quad}||c@{\quad}|c@{\quad}|c@{\quad}|c@{\quad}|c@{\quad}|c@{\quad}|}
  &  $\chi^2/\mbox{n.o.p}$ &  $k_0^2$
($\times 10^{-3}$) &  $\lambda$ &  $R$(GeV${}^{-1}$) &  $\chi^{\prime\prime}(\gamma_c)$ &  $D$ ($\times 10^{-2}$) \\ [0.5ex] \hline \hline
 $\tilde{T}_Y^{\rm{AGBS}}$        &  0.949 &  $3.79 \pm 0.30$ &  $0.213 \pm 0.003$ &  $3.576 \pm 0.059$ &  $4.69 \pm 0.23$ &  $0$ \\ \hline
 $\avg{\tilde{T}_Y^{\rm{AGBS}}}$  &  0.949 &  $3.79 \pm 0.30$ &  $0.213 \pm 0.003$ &  $3.576 \pm 0.059$ &  $4.69 \pm 0.23$ &  $0.0 \pm 1.1$ \\ \hline
\end{tabular}
\caption{Parameters extracted from the fit to $F_2$ H1 and ZEUS data \cite{zeus,h1} in the case where $m_{u,d,s}=50$
MeV.}\label{tab:param50}
\end{table}

Figures \ref{Fig-1} and \ref{Fig-2} show the $F_2$ structure function in bins of $Q^2$ for small and moderate values of $Q^2$, respectively. As usual, the H1 data have been rescaled
by a factor 1.05 which is within the normalization uncertainty. Tables
\ref{tab:param50} and \ref{tab:param140} show the values of the parameters
obtained from the fit both with and without fluctuations. The latter
corresponds to the value $D=0$ for the diffusion coefficient.

\begin{table}
\begin{tabular}{l@{\quad}||c@{\quad}|c@{\quad}|c@{\quad}|c@{\quad}|c@{\quad}|c@{\quad}|}
  &  $\chi^2/\mbox{n.o.p}$ &  $k_0^2$
($\times 10^{-3}$) &  $\lambda$ &  $R$(GeV${}^{-1}$) &  $\chi^{\prime\prime}(\gamma_c)$ &  $D$ ($\times 10^{-3}$) \\ [0.5ex] \hline \hline
 $\tilde{T}_Y^{\rm{AGBS}}$        &  0.942 &  $1.69 \pm 0.16$ &  $0.176 \pm 0.004$ &  $4.83 \pm 0.12$ &  $6.43 \pm 0.29$ &  $0$ \\ \hline
 $\avg{\tilde{T}_Y^{\rm{AGBS}}}$  &  0.942 &  $1.69 \pm 0.16$ &  $0.176 \pm 0.004$ &  $4.83 \pm 0.12$ &  $6.43 \pm 0.29$ &  $0.0 \pm 9.6$ \\ \hline
\end{tabular}
\caption{Parameters extracted from the fit to $F_2$ H1 and ZEUS data \cite{zeus,h1} in the case where $m_{u,d,s}=140$
MeV.}\label{tab:param140}
\end{table}

\begin{figure}[h!]
\input{q2-baixo.tex}
\caption{Predictions for the H1 \cite{h1} and ZEUS \cite{zeus} data for the proton structure function versus $x$ for small values of $Q^2$, given in GeV{}$^2$. The fit was performed with quark masses $m_{u,d,s} = 140$ MeV.}
\label{Fig-1}
\end{figure}

\begin{figure}[h!]
\input{q2-alto.tex}
\caption{Predictions for the H1 \cite{h1} and ZEUS \cite{zeus} data for the proton structure function versus $x$ for moderate values of $Q^2$, given in GeV{}$^2$. The fit was performed with quark masses $m_{u,d,s} = 140$ MeV.}
\label{Fig-2}
\end{figure}




\section{Conclusions and discussion}
In this paper, the AGBS model is used to investigate the possible effects of the gluon
number fluctuations in the HERA data. The expression for the amplitude, Eq.(\ref{eq:Tmodel}),
is considered as a single-event amplitude and the average amplitude is evaluated through
Eq.(\ref{eq:agbs-fluct}) and put into the expression for the proton structure function,
whose data was successfully reproduced. This is shown by the good $\chi^2/\rm{n.o.p.}$
and by the curves in Figures \ref{Fig-1} and \ref{Fig-2}. From the comparison between the
 results with and without fluctuations (see Tables \ref{tab:param50} and \ref{tab:param140})
 one sees that the value of the $\chi^2/\rm{n.o.p.}$ does not change, and the same can be
 said for the parameters. Specially for the diffusion coefficient $D$, its value obtained
 in the case with fluctuations is very small, actually very near its mean field value $D=0$.
 Then, in the framework of AGBS model, there is no evidence of fluctuations in DIS experiment
 at HERA energies. This indicates that a mean field treatment, with fixed coupling, is enough to
investigate high energy QCD phenomenology, at least at HERA energies.

It is interesting to compare our results with those obtained in Ref.\cite{Kozlov:2007wm},
where GBW and IIM models are used to parametrize the average dipole-proton scattering
 amplitude. Although this analysis results in a successful description of HERA data,
it is not conclusive concerning the presence of fluctuations. In particular, the value
 of $D$ extracted from the fit agrees with the values predicted in the literature
 \cite{soyez-fluct,onedim}, i.e., a sizable number of order ${\cal O}(1)$, which would
 indicate that fluctuations could be present at HERA. However, one should point out
 that in \cite{Kozlov:2007wm}, the fit is performed using only data from ZEUS Collaboration
 \cite{zeus}, i.e., it does not include H1 data \cite{h1}, and within a more restricted
 kinematical range, with values of virtuality such that $Q^2_{max}=50$ GeV${}^2$. 

\begin{table}[h!]
\begin{tabular}{l@{\quad}||c@{\quad}|c@{\quad}|c@{\quad}|c@{\quad}|c@{\quad}|c@{\quad}|}
  &  $\chi^2/\mbox{n.o.p}$ &  $k_0^2$
($\times 10^{-3}$) &  $\lambda$ &  $R$(GeV${}^{-1}$) &  $\chi^{\prime\prime}(\gamma_c)$ &  $D$ \\ [0.5ex] \hline \hline
 $\tilde{T}_Y^{\rm{AGBS}}$  &  0.788 &  $4.258 \pm 0.425$ &  $0.214 \pm 0.005$ &  $3.497 \pm 0.068$ &  $4.336 \pm 0.281$ &  $0$ \\ \hline
 $\avg{\tilde{T}_Y^{\rm{AGBS}}}$ & 0.782 & $4.023 \pm 0.560$ & $0.190 \pm 0.030$ & $3.644 \pm 0.214$ & $3.840 \pm 0.214$ &  $0.922 \pm 1.162$ \\ \hline 
\end{tabular}
\caption{Parameters extracted from the fit to $F_2$ ZEUS data \cite{zeus} in the case where $m_{u,d,s}=50$ MeV.}\label{table:param-zeus:50}
\end{table}

\begin{table}[h!]
\begin{tabular}{l@{\quad}||c@{\quad}|c@{\quad}|c@{\quad}|c@{\quad}|c@{\quad}|c@{\quad}|}
  &  $\chi^2/\mbox{n.o.p}$ &  $k_0^2$
($\times 10^{-3}$) &  $\lambda$ &  $R$(GeV${}^{-1}$) &  $\chi^{\prime\prime}(\gamma_c)$ &  $D$  \\ [0.5ex] \hline \hline
 $\tilde{T}_Y^{\rm{AGBS}}$ &  0.778 &  $1.965 \pm 0.222$ &  $0.177 \pm 0.006$ &  $4.681 \pm 0.136$ &  $5.946 \pm 0.944$ & $0$ \\ \hline
$\avg{\tilde{T}_Y^{\rm{AGBS}}}$ & 0.768 & $1.383 \pm 0.118$ & $0.120 \pm 0.010$ & $5.459 \pm 0.043$ & $5.464 \pm 0.547$ &  $1.778 \pm 0.381$ \\ \hline 
\end{tabular}
\caption{Parameters extracted from the fit to $F_2$ ZEUS data \cite{zeus} in the case where $m_{u,d,s}=140$ MeV.}\label{table:param-zeus:140}
\end{table}

Actually, if one performs the same analysis of Ref.\cite{Kozlov:2007wm}--only
ZEUS in the same kinematical range--using the AGBS model, the results obtained are very similar, as it can be seen in Table \ref{table:param-zeus:140}, with a better $\chi^2/\rm{n.o.p.}$. The present analysis is then more complete, since it includes all the last HERA data and considers a wider kinematical range.

Our conclusions seem to shed some light on the investigation of the effects
of fluctuations at HERA and confirm the robustness of the AGBS model. Of course,
only in the near future, at LHC, it will be possible to see if fluctuations
are present at much higher energies, or if they are really suppressed by
the running of the coupling, as it has been suggested by recent developments on a one-dimensional toy model which successfully reproduces the main features
of scattering and high energy evolution in QCD \cite{onedim,Dumitru:2007ew}.

\section*{Acknowledgements}
This work is partially supported by CNPq.

\bibliographystyle{unsrt}
 
\end{document}